\documentstyle[psfig]{article}

\begin{document}

\begin{center}
{\LARGE Discovery of Non-Thermal X-Rays from the Shell of RCW~86}\\
\vspace{0.6cm}
Aya {\sc Bamba}, Katsuji {\sc Koyama}\footnote{CREST, Japan Science and 
   Technology Corporation (JST), 
   4-1-8 Honmachi, Kawaguchi, Saitama 332-0012 JAPAN} \\
{\it Department of Physics, Graduate School of Science, Kyoto University, 
Sakyo-ku, Kyoto 606-8502} \\
{\it E-mail(AB): bamba@cr.scphys.kyoto-u.ac.jp} \\
and \\
Hiroshi {\sc Tomida} \\
{\it National Space Development Agency of Japan,} \\
{\it World Trade Center Bldg,
2-4-1 Hamamatu-cho,Minato-ku, Tokyo,105-8060}
\end{center}
We report the ASCA (Advanced Satellite for Cosmology and Astrophysics) results
of RCW~86, a shell-like supernova remnant (SNR).
The bright region in the X-ray band traces the radio clumpy shell, 
although details of the structure are different.
The X-ray spectrum from each part of the shell can not be fitted to
a thin thermal plasma model, 
but requires, at least three components:
a low temperature plasma of 0.3~keV, high temperature plasma of
$\geq $ several keV,
and a power-law component with a photon index $\sim\ 3$.
The abundances of O, Ne, Mg and  Si are significantly higher than that of Fe, 
indicating that RCW~86 is a type~II SNR.
The absorption column of $\sim ~3\times ~10^{21}~{\rm H}~{\rm cm}^{-2}$ 
indicates the distance to the SNR to be several kpc.
The power-law component can be interpreted to be synchrotron radiation of
high energy electrons.
Assuming energy density equipartition
between the magnetic field and the electrons,
and using the radio and X-ray spectra,
we argue that high energy electrons are accelerated up to 20~TeV.
The acceleration efficiency is, however,
different from shell to shell.


\section{Introduction}

The origin and acceleration of cosmic rays have been
a long-standing key problem since the discovery in 1911-1912
(Hess 1911; Hess 1912).  
The energy spectrum of cosmic rays is presented by a  power-law function
with a break at $\sim 10^{15.5}$~eV called ``knee energy''.
The presence of the break in the power-law distribution indicates 
that cosmic rays may have two different origins.
A natural explanation is that cosmic rays below the knee energy are
originated in our Galaxy,
while those above the knee are extragalactic origin. 
The most probable origin and acceleration mechanism 
of the Galactic cosmic rays are the first order Fermi acceleration process 
by fast moving shells of supernova remnants (SNRs) (Wentzel 1974).
In fact, radio spectra and polarizations in SNR shells are
well represented by synchrotron radiations of high energy electrons
with a power-law distribution of an index expected
from the Fermi acceleration process.
The radio data thus indicate that GeV energy electrons are accelerated
in SNR shells.  

If electrons are accelerated to a higher energy near the knee,
the energy of synchrotron radiation shifts to an X-ray band.
Shell-like SNRs, however, typically display line-dominated thermal X-rays,
with the characteristic temperature of about 1~keV,
due to either a single or multi-components plasma.
Thus no evidence for synchrotron X-rays has been so far found.
Recently two shell-type SNRs,
SN~1006 and G347.3$-$0.5 are found to be dominated by power-law X-rays
with no emission line
(Koyama et al. 1995; Koyama et al. 1997; Slane et al. 1999).
The power-law X-rays, together with the discoveries of TeV gamma rays
(Tanimori et al. 1998, Muraishi et al. 2000),
have indicated that the shell of these SNRs are the acceleration sites of
extremely high energy electrons of 10$^{13}$ to 10$^{15}$ eV.
In such relativistic energy, protons, the main component of cosmic rays,
and electrons are essentially the same except for the charge polarity
(Bell, 1978).
Therefore, these two SNRs are most probable sites of cosmic ray accelerations
near to "knee energy".  

Some hints for synchrotron X-rays from other shell-like SNRs are found
with ASCA (Advanced Satellite of Cosmology and Astrophysics),
RXTE (Rossi X-ray Timing Explorer) and CXO (Chandra X-ray Observatory).
Even typical thin thermal SNRs, such as Cas A, IC443, Tycho, Kepler, 
G156.2$+$5.7 and RCW~86 exhibit hard energy tails
over the thin thermal emissions.
Some of the hard tails are most likely to be non-thermal
(Cas A and IC443; Petre et al. 1997),
but the others are controversial,
either non-thermal or higher energy plasma of about 10 keV.
Since high energy particles from a higher temperature plasma
(e.g.\ $\ge$ 10~keV)
are more easily injected into the "Fermi acceleration machine"
than those from a lower temperature plasma (e.g.\ $\sim$ 1~keV)
(Bell, 1978),
it is essentially important to investigate the relation 
of high energy plasma and non-thermal emission.

RCW~86 was identified as the remnant of the historical supernova in A.D.185
(Clark and Stephenson 1977).
However Chin and Huang (1994), and Schaefer (1995), derive the age of RCW~86
to be $\sim~8000$~yrs with a Sedov solution.
Since RCW~86 is located in an OB star association,
it is likely to be a type II SNR (Clark and Stephenson, 1977).
X-rays from RCW~86 were discovered by Naranan et al. (1977).
The X-ray spectrum obtained with the Einstein satellite was represented by 
a two-temperature plasma model (Winkler 1978).
The higher energy spectrum (2.0--20~keV) with the Ginga satellite was
explained by a single temperature plasma of $kT \ge$ 4~keV
(Kaastra et al.\ 1992).
Using the ASCA satellite, Vink et al.\ (1997) reported that
the plasma temperature varies from region to region of the shells,
from 0.8~keV to $>$~3~keV.
Petre et al.\ (1999) found a hard X-ray tail extending 
to several 10~keV in the RXTE spectrum.
They thus argued that the shell of RCW~86 has a non-thermal emission
and is one of the candidate for the origin of high energy cosmic rays.
However, non-imaging instrument of RXTE could not make clear
which region of the SNR is the hard X-ray emitter.
In this paper, we revisit to the ASCA data and examine
whether or from which region is RCW~86 emitting non-thermal hard X-rays.

\section{Observations}

RCW~86 was observed with ASCA (Tanaka et al.\ 1994) on 1993 August 17-18.
X-ray photons are collected with four XRTs
(X-Ray Telescopes; Serlemitsos et al.\ 1995)
and detected with two SIS (Solid-state Imaging Spectrometers;
Burke et al. 1991) cameras
and two GIS (Gas Imaging Spectrometers; Ohashi et al.\ 1996)
on the foci of the XRTs.
Three mosaic observations have been conducted, because the size of RCW~86
is larger than the field of view of SIS.
The SISs were operated in 4-CCD read-out mode,
with the data acquisition of faint mode in the High bit rate and
bright mode in the Medium and Low bit rates.
No significant degradation is found in the SIS data taken
about 6 months after the launch,
hence no correction for the Residual Dark Distribution (RDD)
is required (T. Dotani et al.\ 1997, ASCA News 5, 14).
The GISs were operated in the normal PH mode.
We excluded high-background data and non X-ray events 
with the standard method according to the user guide
by NASA Goddard Space Fright Center.
Then, the available exposure times are $\sim~20$~ks, $\sim~9$~ks,
and $\sim~8$~ks, for each observation.

\section{Analyses and Results}
\begin{figure}[bhtp]
\begin{tabular}{cc}
\psfig{figure=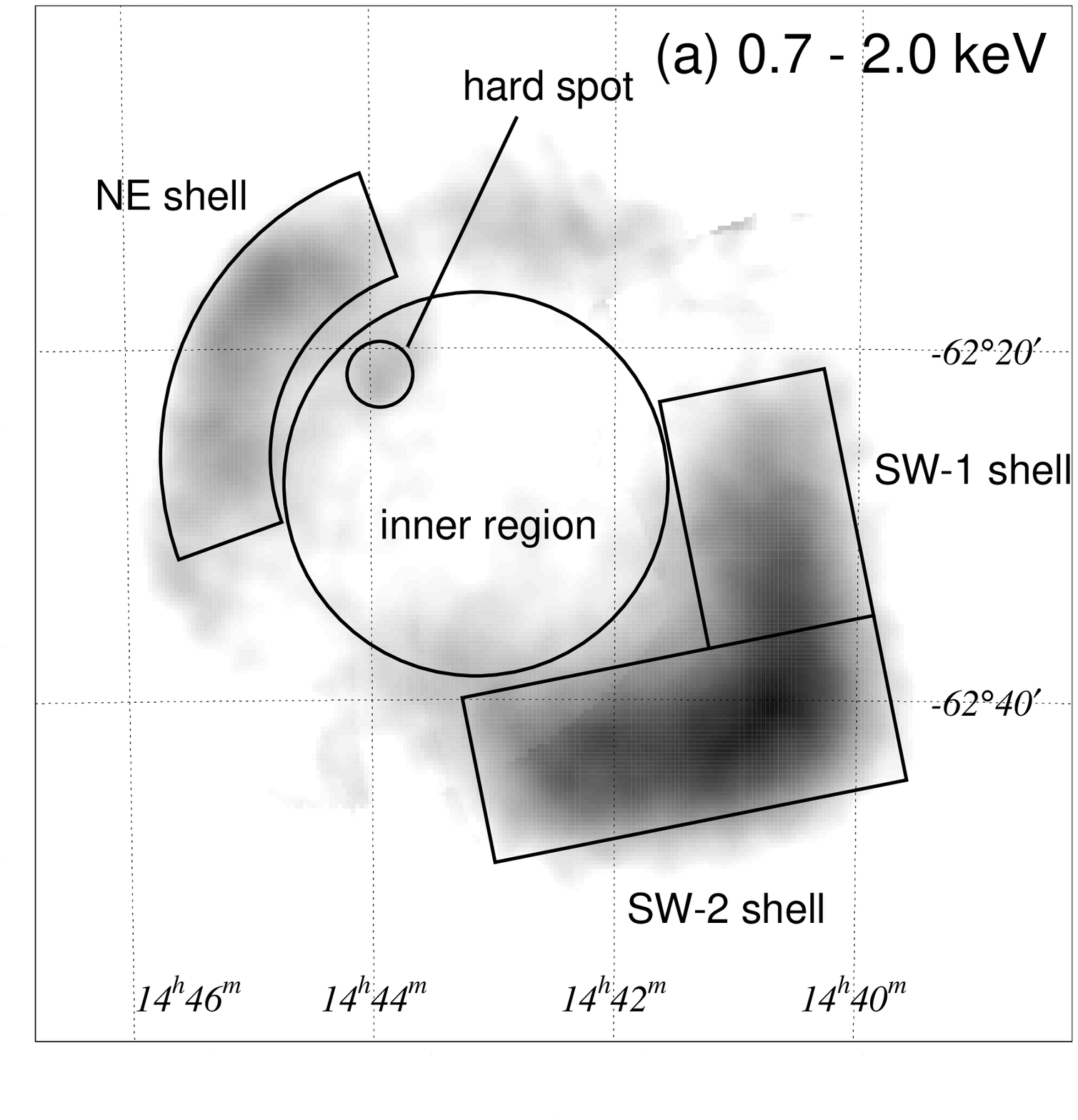,width=6cm} &
\psfig{figure=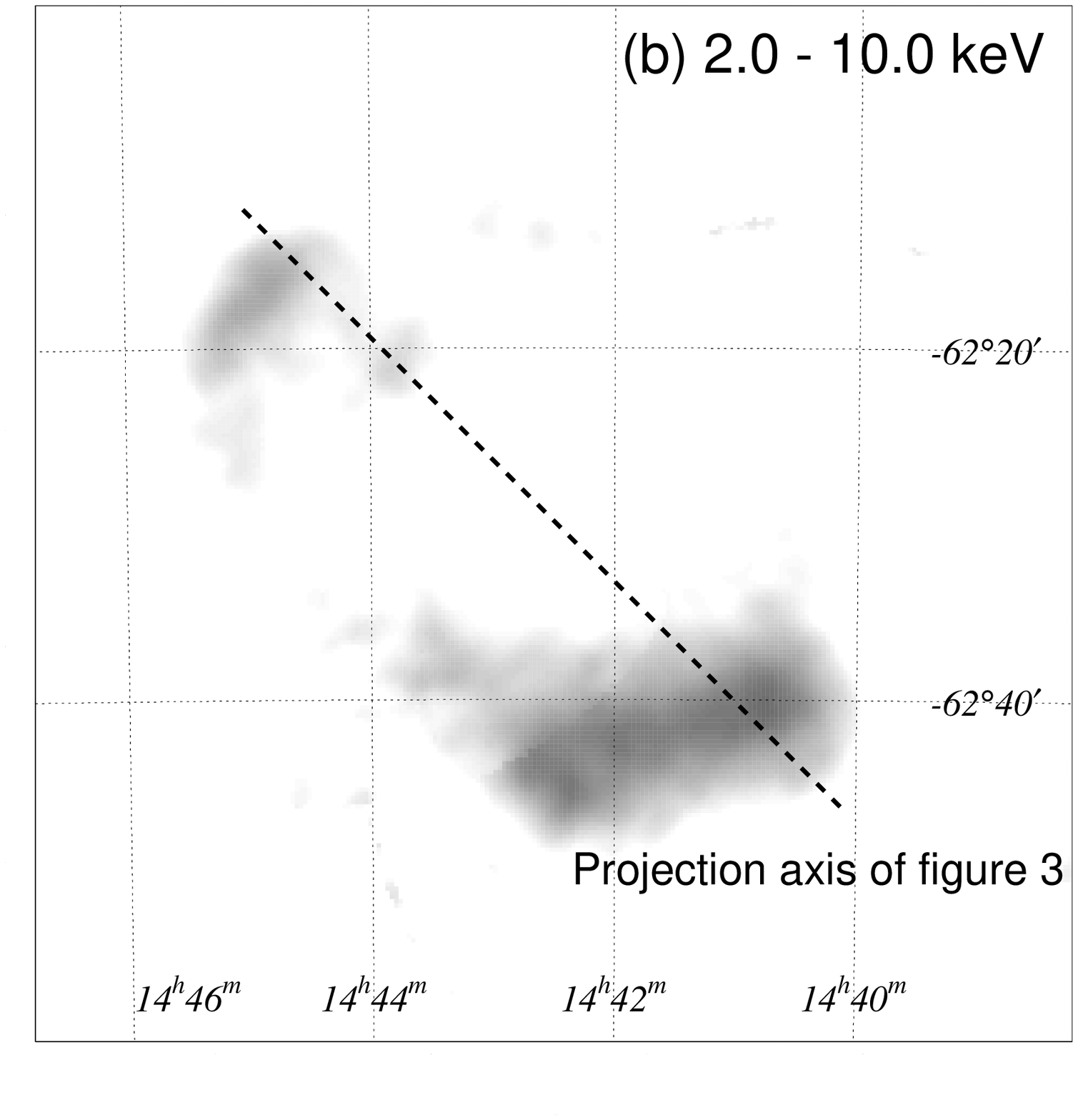,width=6cm}
\end{tabular}
\caption{Mosaic GIS maps in the 0.7--2.0~keV (a) 
and the 2.0--10.0~keV (b) bands with J2000 coordinate.
Vignetting and exposure time have been  corrected.
The gray map is drawn in logarithmic scale.}
\end{figure}

Figure~1 shows the GIS mosaic images around RCW~86 
in the energy bands of 0.7--2.0~keV (figure~1a)
and 2.0--10.0~keV (figure~1b),
in which vignetting and exposure times are corrected.
In both the images, we can see two prominent shells
at northeast and southwest of the RCW~86 rim.
The sizes and shapes of the X-ray shells are similar to those of
H$\alpha$ (Smith 1997) and the 847~MHz band (Kesteven and Caswell 1987).
With a quick look, we see different structure between the two energy bands,
or the spectrum is different from region to region.  
In the 0.7--2.0~keV band, the southwest shell is appeared
to be boomerang shape,
while in the 2.0--10.0~keV band, the shell changes to bar-like structure.
The northeast shell is brighter in the hard band than in the soft band.
We also detected a point-like spot near the northeast shell 
in figure~1a with a significance of 4$\sigma$.

We made the X-ray spectra from each region with the designation given 
in figure~1a.
Since RCW~86 is located near on the Galactic plane,
contamination from the Galactic ridge plasma emission can not be ignored (Koyama et al. 1987).
Hence the background spectrum is taken from outside of RCW~86
but near the same Galactic latitude as the source region,
and is subtracted from the source spectrum.  
We first made the spectra from the brightest shell, SW shell
(combined with SW-1 and SW-2 shells).
Although we treated both the GIS and SIS spectra simultaneously
to increase statistics, we show, for simplicity, only the SIS spectrum in figure~2a.
Prominent emission lines from highly ionized O, Ne, Mg, Si and Fe atoms
in the spectrum imply that significant fraction of the X-rays
is attributable to a thin thermal plasma.
Therefore, we applied a thin thermal model in non-equilibrium ionization
(Masai 1984) as was already tried by Vink et al. (1997).
This model is statistically rejected with  $\chi^2$ = 2154/685,
even the abundances of O, Ne, Mg, Si and Fe are allowed to be free parameters.
Large residuals are found in the 1--2~keV and above 5~keV bands.
We note that the thermal model reported by Vink et al. (1997)
is unacceptable to our GIS and SIS simultaneous fitting.
In fact, the iron K-shell line and those from lighter elements
such as  O, Ne, Mg and Si can hardly coexist in a single temperature plasma. 
We thus added another higher temperature thin plasma
in non-equilibrium ionization.
This two-temperature plasma model is still rejected
with $\chi^2$ = 1685/682.
Significant data excess is found at high energy continuum above 6.5~keV.
The discrepancy of the model and data become larger in higher energy.
In fact, this model spectrum can only account 10\% of the RXTE flux
at $>$~10~keV reported by Petre et al. (1999).

\begin{figure}[hbtp]
\psfig{figure=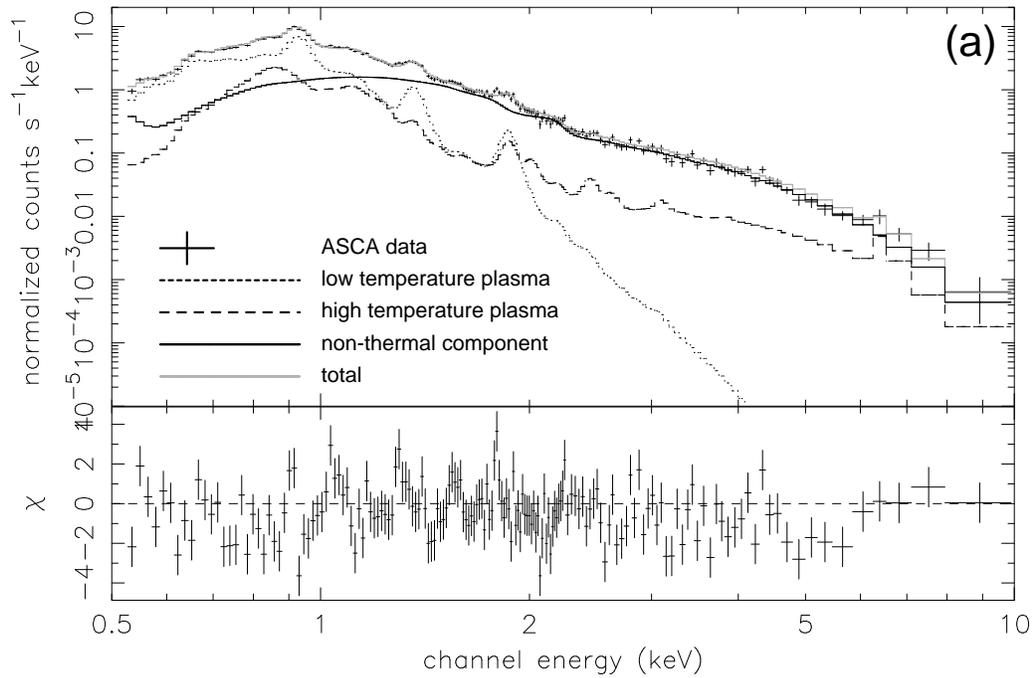,width=15cm}
\psfig{figure=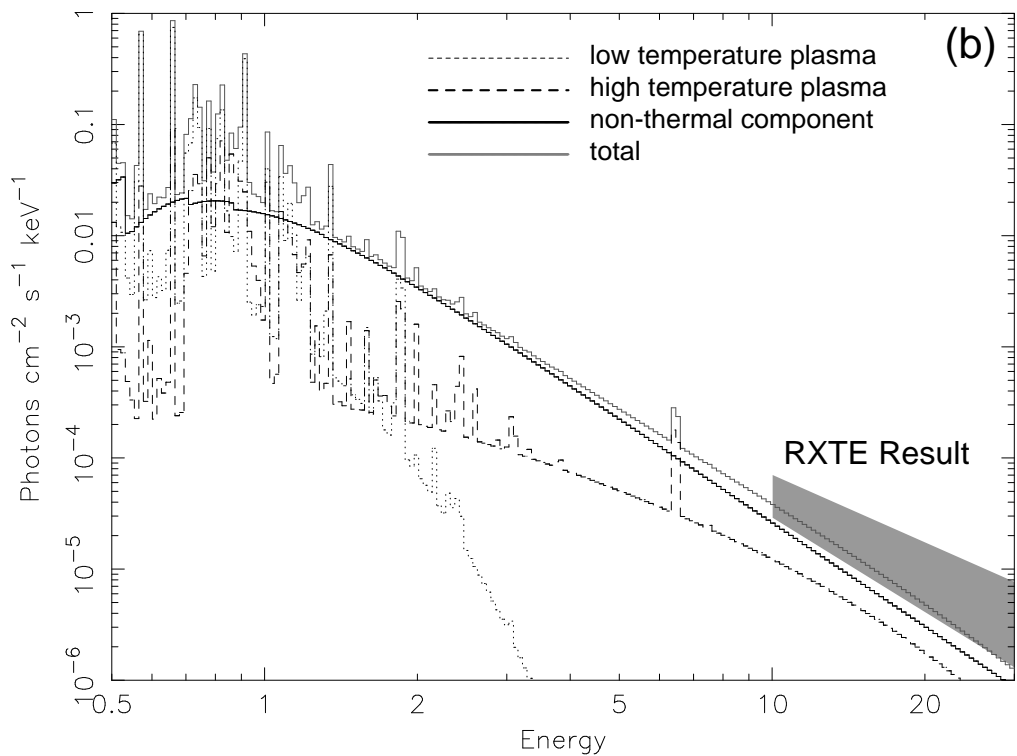,width=15cm}
\caption{ (a) The ASCA SIS~0+1 spectra from SW shell of RCW~86  with the best-fit models (upper panel).
Residuals from the best-fit models are shown in the lower panel. 
The dotted, dashed, and solid line represents the best-fit models(see text).
(b) The photon flux of the best-fit model. The hatched area represents the photon 
fluxes and errors with the RXTE observations.}
\end{figure}

Accordingly, we are required to add a non-thermal (power-law) component,
hence applied a model consisting with two thin thermal plasmas
component~1 (dotted line) and 2 (dashed line), respectively,
and a power-law component (solid line, see figure~2a).
For simplicity, the absorption column is taken to be common
among the three components,
and the abundances of O, Ne, Mg, Si, and Fe in component~1 and
that of Fe in component~2 are allowed to be free parameters,
while the others are fixed to be solar values (Allen 1973).
$\chi^2$ is then greatly reduced to be marginally value of 979/682.
Therefore we do not try more complicated models,
although this 3-component model is still rejected
in a statistical point of view.
The best-fit parameters are shown in table~1.

\begin{table*}[t]
{\small
\begin{center}
Table~1.\hspace{4pt}Best-fit parameters for SW shell for a
two-temperature plasma and power-law model.$^\ast$\label{tbl-1}\\
\end{center}
\vspace{6pt}
\begin{tabular*}{\textwidth}{@{\hspace{\tabcolsep}
\extracolsep{\fill}}p{12pc}ccc}\hline\hline\\ [-6pt]
Parameters & Component~1 & Component~2 & Non-thermal component
\\[4pt]\hline\\[-6pt]
$kT$ (keV)\dotfill & 0.34 (0.30 -- 0.38) & 23.8 (15.1 -- 26.0) & $\cdots$ \\
$\log nt$ (s cm$^{-3}$)\dotfill & 10.7 (10.6 -- 10.8) & 10.1 (10.0 -- 10.2) 
	& $\cdots$ \\
O/H$^\dagger$\dotfill & 3.3 (2.3 -- 3.8) & $\cdots$ & $\cdots$ \\
Ne/H$^\dagger$\dotfill & ($>$ 6.7) & $\cdots$ & $\cdots$ \\
Mg/H$^\dagger$\dotfill & ($>$ 4.7) & $\cdots$ & $\cdots$ \\
Si/H$^\dagger$\dotfill & ($>$ 6.5) & $\cdots$ & $\cdots$ \\
Fe/H$^\dagger$\dotfill & 2.4 (1.7 -- 3.1) & 3.3 (2.3 -- 6.0) 
& $\cdots$ \\
photon index\dotfill & $\cdots$ & $\cdots$ & 3.1 (3.0 -- 3.2) \\
$N_{\rm H}\ (\times 10^{21}\ {\rm H}\ {\rm cm}^{-2})$\dotfill 
& 3.2 (2.9 -- 3.8) & 
3.2$^\ddagger$ & 3.2$^\ddagger$ \\
Flux (erg cm$^{-2}$ s$^{-1}$)$^{\S}$\dotfill 
& 1.5E$-$10 & 6.2E$-$11 & 1.0E$-$10 \\ \hline
\end{tabular*}
\vspace{6pt}\par\noindent
$^\ast$ Parentheses indicate 90~\% confidence regions for 
one relevant parameter.
\par\noindent
$^\dagger$ Abundance ratio relative to the solar value.
\par\noindent
$^\ddagger$ Common with component~1.
\par\noindent
$^{\S}$ In the 0.7--10.0~keV band.}
\end{table*}

 The power-law fluxes extrapolated to higher than 10~keV
are smoothly connected to the RXTE results as is shown in the
best-fit model function given in figure~2b.
Thus the hard X-rays obserevd with RXTE can be regarded to be
a higher energy part of the power-law component
found with the present ASCA analysis of the brightest shell.

For the diffuse spectra from the other regions of this SNR,
we applied the same model as SW shell,
allowing only the normalizations of the three components
and the ionization parameter to be free,
otherwise the limited photon statistics of these faint regions
give no essential constraint on the physical parameters.
This simplified model approach is found to be reasonable,
because the reduced $\chi^2$ for the fitting of each region is about 1.5--1.8,
comparable to that in the SW shell analysis.
The best-fit ionization parameters, fluxes and the flux ratios
for each component are shown in table~2.

\begin{figure}[hbtp]
\psfig{figure=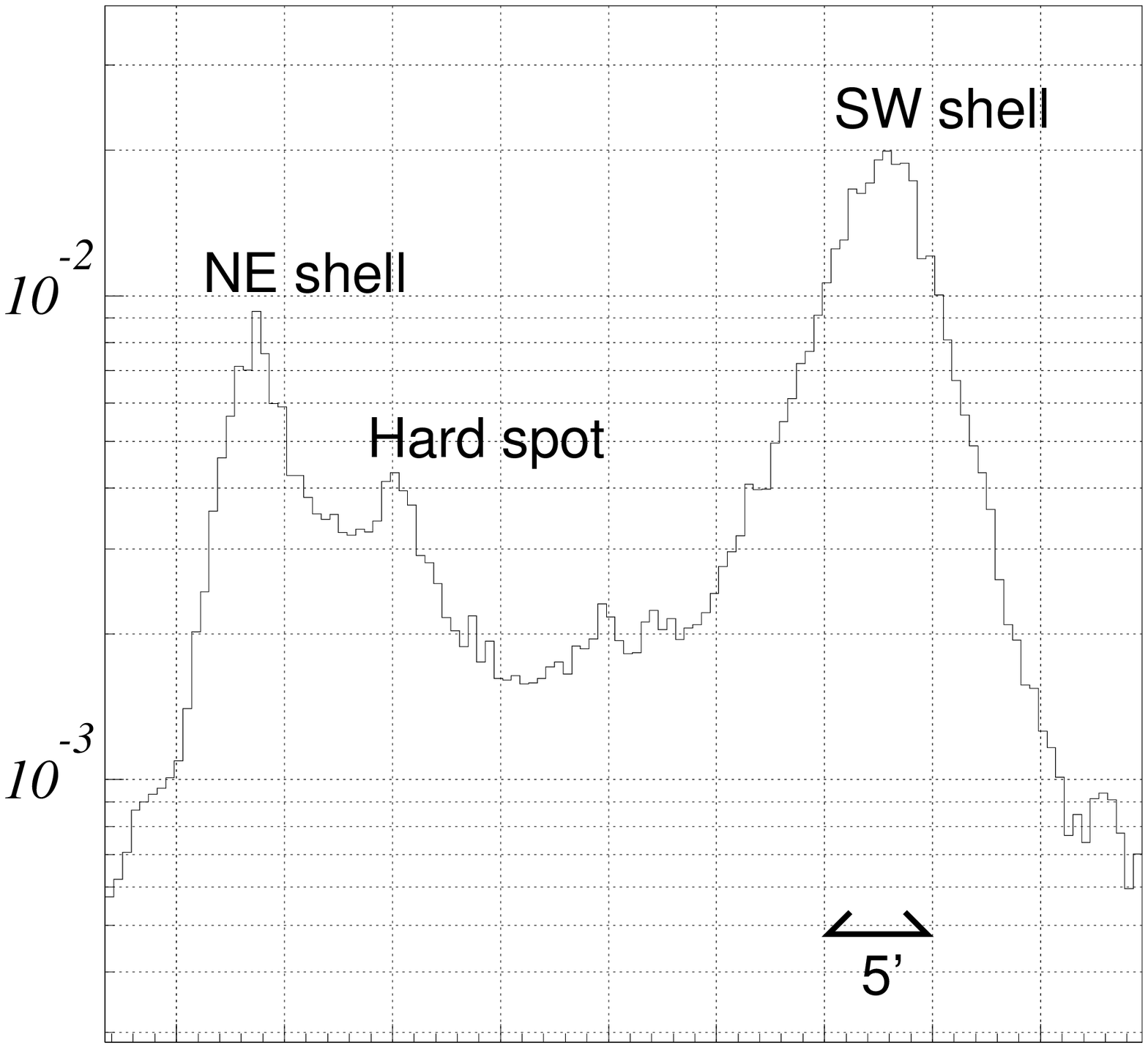,width=8cm}
\caption{The projected profile to the axis given in figure~2b).
The vertical axis represents count rate per bin.}
\end{figure}

Figure~3 is a projected profile of RCW~86 to the dotted line given in figure~2b.
We can clearly see a hard spot near the NE shell, with the width of about 3$^\prime$, 
corresponding to the spatial resolution of GIS.
We thus regard the hard spot to be a point source.
The spectrum of the hard spot is taken from
a circle of $3^\prime$ radius on the source,
and that of the background is an annulus region of
$3^\prime$--$6^\prime$ radius around the source.
The background subtracted spectrum shows no significant line,
and can be fitted to a power-law model
of a photon index~$ = 2.3^{+0.2}_{-0.2}$
and absorption of $N_{\rm H} = 2.0^{+0.9}_{-0.8}\times 10^{21} {\rm cm}^{-2}$
with reduced $\chi^2$ = 51/70
(here and after, errors are 90~\% confidence).
The total flux in the 0.7--10.0~keV band is
$3.7\times 10^{-12} {\rm erg\ cm}^{-2}{\rm s}^{-1}$.  

\section{Discussions}

We found that the X-ray spectra of RCW~86 have three components;
a lower temperature plasma (component 1),
higher temperature plasma (component 2),
and a power-law emission (non-thermal component).
Since the abundance of light elements such as O, Ne, Mg, and Si
are higher than that of Fe,
RCW~86 is likely due to a type~II SN,
consistent with that RCW~86 is in an OB star association
(Clark and Stephenson 1977).
The best-fit column density of $3.2\times 10^{21}\ {\rm H\ cm}^{-2}$
is also consistent with the distance to
the OB star association of $2.5~{\rm kpc}$.
Here and after, we adopted this value for the distance to RCW~86.
The discovery of the hard spot in RCW~86 is also suggestive
for the Type II concept.
We separately discuss for these emissions.

\subsection{Hard Spot}

Since the absorption column of the hard spot is similar to that of SW shell,
we infer that the hard spot is located in the SNR.
The total luminosity is then estimated
to be $2.7~\times~10^{33}\ {\rm erg\ s}^{-1}$.
This luminosity and the power-law spectrum of a photon index ($\alpha = 2.3$)
suggest the hard spot could be a Crab-like pulsar,
supporting for a type~II SN origin.
We therefore search for coherent pulsations using the FFT technique,
however we found no hint of pulsation in the time scale of
7.825$\times 10^{-3}$~sec to 16381~sec.
Obviously, further sensitive searches for coherent pulsations are required.

\subsection{Higher Temperature Plasma}

The nature of component~2 is puzzling because the temperature of 
$\ge$~10~{\rm keV} is unusually high for SNRs.
The presence of this high temperature plasma mainly rely on
the line features around 6.5~keV,
because the continuum emissions above a few keV are not sensitive
to separate a high temperature plasma and a power-law emission.
Thus to investigate the origin of the line feature is essentially important.
Since the observed line energy is about 6.5~keV,
near the K-shell transition energy of neutral or low ionization iron atoms,
one may argue that the line is fluorescence from cold iron, like
in a molecular cloud Sgr~B2 (Murakami et al.\ 2000).
This possibility can be rejected because the column density is too small
to emits such strong neutral iron lines by the reflection mechanism.
Strong iron lines from rather low ionization atoms are also found
in other young SNRs, for example, Tycho and Kepler's SNRs
(Hwang and Gotthelf 1996).
Borkowski and Szymkowiak (1997) suspected that these iron lines in young SNRs
are the composite of lines from a hot plasma and fluorescent lines
coming from dust grains.
If this is the case of RCW~86,
the energy of iron line due to a hot plasma can be higher,
then the ionization parameter become larger.
Hence the true temperature of the hot plasma should be lower than 10~keV,
probably around several keV, which is not unusually high for a young SNR.   

\subsection{Lower Temperature plasma}

At the distance of 2.5~kpc,
the radius of the shell $(R)$ is estimated to be 14~pc.
Assumption that SW shell covers $\pi /2$ str (see figure~1)
and that the thickness is $1/12R$ (expected from Sedov solution),
the density and ionization age of SW shell are estimated 
from the luminosity and the ionization parameter $nt$,
to be 0.45 (0.40--0.50)~cm$^{-3}$ and 3100~(2500--3700)~yr, respectively.
For NE shell, we assume the covering solid angle to be $\pi /4$ str
(see figure~1)
and estimate the density and ionization age to be 0.06~(0.04--0.08)~cm$^{-3}$
and 1800 (1000--3900)~yr, respectively.
These ionization ages are significantly shorter than the Sedov age of 8000~yr
(Chin and Huang, 1994; Schaefer, 1995),
and favor that RCW~86 is the remnant of AD~185 (Clark and Stephenson 1977).
We note that systematical difference between the ionization age
and the Sedov age is found in some other Type II SNRs
by Hughes et al.\ (1998).
They have systematically analyzed shell-like SNRs
in the Large Magellanic Cloud
and found some of the Type II candidates shows the ionization ages
shorter than the Sedov ages.
They interpret that these Type II SNRs exploded within pre-existing
low-density cavities in the interstellar medium.
The low interior density allows the SN blast wave to propagate rapidly
to the cavity wall,
where it then encounters denser gas, begins to slow down,
and emits copious amounts of X-rays.
This scenario reduces the dynamical ages from that inferred
with the Sedov model (Sedov ages).
RCW~86 may be the same case,
because it is likely a Type II in an OB star association
(Clark and Stephenson 1977).
Thus the real age would be similar to the ionization age of 3100~yr.

\subsection{Power-law Emission}

\begin{figure}[hbtp]
\psfig{figure=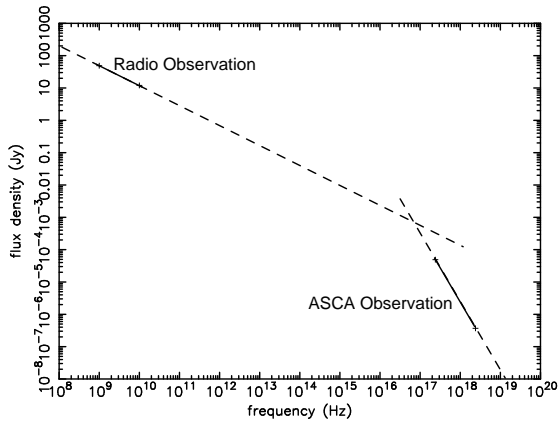,width=8cm}
\caption{The wide band spectrum from radio to X-ray band.
The solid lines mean observational flux density, 
whereas the dashed lines are extrapolated flux.}
\end{figure}

Figure~4 is the wide band spectrum from RCW~86, 
in which the radio data is refereed from Green (1998).
The spectrum can be explained by a power-law model with a break
at $\sim 2\times 10^2\ {\rm eV}$.
This break is produced by synchrotron energy loss,
which is larger for higher energy electrons,
hence gives the maximum electron energy.
The synchrotron energy loss is proportional to
the square of the magnetic field.
The magnetic field is calculated using the IRAS result
by Greidanus and Storm (1990) to be $80~\mu {\rm G}$
under the assumption of energy equipartition.
We then estimate the highest energy of electrons which have been accelerated
in the shell of RCW~86 to be $\sim\ 20~{\rm TeV}$.
Actually, many SNRs have smaller magnetic field
than that on the assumption of energy equipartition.
Therefore, the highest energy of electrons may be higher than our estimation.

\begin{table*}[bt]
\begin{center}
Table~2.\hspace{4pt}Normalized flux of each region 
for the thermal and non-thermal components.$^\ast$ \label{tbl-2}
\end{center}
\begin{tabular*}{\textwidth}{@{\hspace{\tabcolsep}
\extracolsep{\fill}}p{6pc}ccccc}\hline\hline\\ [-6pt]
Region & \multicolumn{2}{c}{thermal component} & 
non-thermal component & N/T$^\dagger$ &
reduced $\chi^2$ \\
 & \multicolumn{2}{c}{(erg\ s$^{-1}$\ cm$^{-2}$\ arcmin$^{-2}$)}
 & (erg\ s$^{-1}$\ cm$^{-2}$\ arcmin$^{-2}$) & & \\
 & component~1 & component~2 & & \\[4pt]\hline\\[-6pt]
SW-1 shell\dotfill & 6.9E$-$13 & 1.6E$-$13 & 8.9E$-$14 & 0.1 & 520.5905/291 \\
SW-2 shell\dotfill & 5.1E$-$13 & 2.3E$-$13 & 4.2E$-$13 & 0.6 & 1056.154/673 \\
NE shell\dotfill & 2.5E$-$14 & 3.1E$-$14 & 1.8E$-$13 & 3.2 & 409.3046/256 \\
Inner region\dotfill & 3.2E$-$16 & 4.3E$-$14 &
 5.8E$-$14 & 1.3 & 135.1856/85 \\ \hline
\end{tabular*}
\vspace{6pt}\par\noindent
$^\ast$ In the 0.7--10.0~keV band.
\par\noindent
$^\dagger$ Flux ratio of the non-thermal to the thermal
(component~1 + component~2).
\end{table*}

The flux ratio of the non-thermal to thermal components is different
from region to region.
The thermal component is dominant in SW-1 shell,
while SW-2 shell has both components.
In NE shell, the non-thermal component is dominated.
Both components are dim in the inner region,
supporting that the non-thermal X-rays do not come from a pulsar nebula.
These differences would be related to the difference of the shock intensities,
the magnetic fields, and/or field line direction relative
to that of the shock propagation.
To address these issues,
we need much higher spatial resolution instruments in the hard energy band,
such as Chandra and Newton.\par

\vspace{1pc}\par
{\it Note added in proof}

After completing this draft, we have become aware of a pre-print
by Borkowski et al.
(astro-ph/0006149), which gives essentially the same conclusion
with different approach of data analysis.

\vspace{1pc}\par
We are grateful to all members of the ASCA team.
Our particular thanks are due to K. Yoshita for fruitful 
comments and suggestion about NEI models.

\end{document}